\documentclass[sigconf, nonacm]{acmart}
\AtBeginDocument{%
  }

\usepackage{graphicx}
\usepackage{tabularx}

\begin{document}

\title{Adapting for AI: How elementary teachers adjust their practices for an AI-integrated curriculum}

\author{Fasika Melese}
\affiliation{%
  \institution{University of Pennsylvania}
  \city{Philadelphia}
  \country{United States}
}
  \email{fasikaye@upenn.edu}

\author{Ruiyang Wu}
\affiliation{%
  \institution{University of Pennsylvania}
  \city{Philadelphia}
  \country{United States}
}

  \email{ruiyangedu@gmail.com}

\author{Xinyue Cui}
\affiliation{%
  \institution{University of Florida}
  \city{Gainesville}
  \country{United States}
}
  \email{cuixinyue@ufl.edu}

\author{Joanna Perkins}
\affiliation{%
  \institution{Montgomery County Schools}
  \city{Troy}
  \country{United States}
}
  \email{Joannacperkins@gmail.com}

\author{Xiaoyi Tian}
\affiliation{%
  \institution{Kennesaw State University}
  \city{Marietta}
  \country{United States}
}
  \email{xtian5@kennesaw.edu}
  
\author{Tiffany Barnes}
\affiliation{%
  \institution{North Carolina State University}
  \city{Raleigh}
  \country{United States}
}
  \email{tmbarnes@ncsu.edu}
  
\author{Shiyan Jiang}
\affiliation{%
  \institution{University of Pennsylvania}
  \city{Philadelphia}
  \country{United States}
}
  \email{jiang33@upenn.edu}


\begin{abstract}
Conversational AI tools are entering children's everyday experiences, and schools are interested in adopting them. However, successful classroom integration depends not only on the technology but also on the work teachers do to make it usable and appropriate for their students and classroom context. There is little known about how elementary teachers work as they implement conversational AI tools in real classrooms. In this study, we examine three teachers' experiences implementing an AI literacy and English Language Arts (ELA) curriculum built around ToyTalk, a conversational AI toy development platform, over 13 instructional days, a three-week summer camp. Drawing on daily individual reflections, group reflections, and post-camp interviews, we find that teachers' adaptive practices of repair, differentiation, translation, and balancing sit at the intersection of three tensions (technology, learner, and instruction). Teachers' understanding of AI and their role evolved over the camp experiences. From these findings, we contribute design implications and considerations for deploying conversational AI within elementary classrooms. 
\end{abstract}



\maketitle

\section{Introduction}
Conversational agents and AI-powered toys that reshape what teachers are asked to deliver are arriving in elementary classrooms \cite{akgun_artificial_2022}. Much of the attention has centered on what these systems do for students in terms of engagement, learning gains, or children's conception of AI \cite{yim_artificial_2025}. Yet whether such technologies work in a real classroom depends heavily on the teacher, in a complex classroom setting in real time. Students in the elementary grade levels are still developing reading, writing and basic computer skills that AI activities assume. In addition, most AI literacy curricula are designed for secondary and higher education levels. Therefore, elementary teachers carry a duty of care and various ways of navigating these challenges when interacting with their students to teach AI literacy. Therefore, how teachers make sense of and adapt to these student-facing AI tools is crucial to the success of any intervention in their classroom. 

Teaching is a situated action. Teachers continuously response to what a specific classroom needs in the moment instead of just executing prescribed materials. Prescriptive designs and lived classroom practice are thus in persistent tension \cite{lampert_how_1985}. When the prescribed material involves a classroom tool that is a conversational toy that can behave unpredictably this gap between design and implementation widens. Prior work on human-AI co-orchestration has examined how teachers share control with AI systems \cite{du_boulay_human-ai_2023, yang_pair-up_2023}. However, they focus on teacher-facing analytics and tutoring tools with older students. There is a gap in understanding about how teachers implement student-facing conversational toys with elementary-level students, and how their role evolves over the course of time. 

\begin{figure}[htbp]
    \centering
    \includegraphics[width=\linewidth]{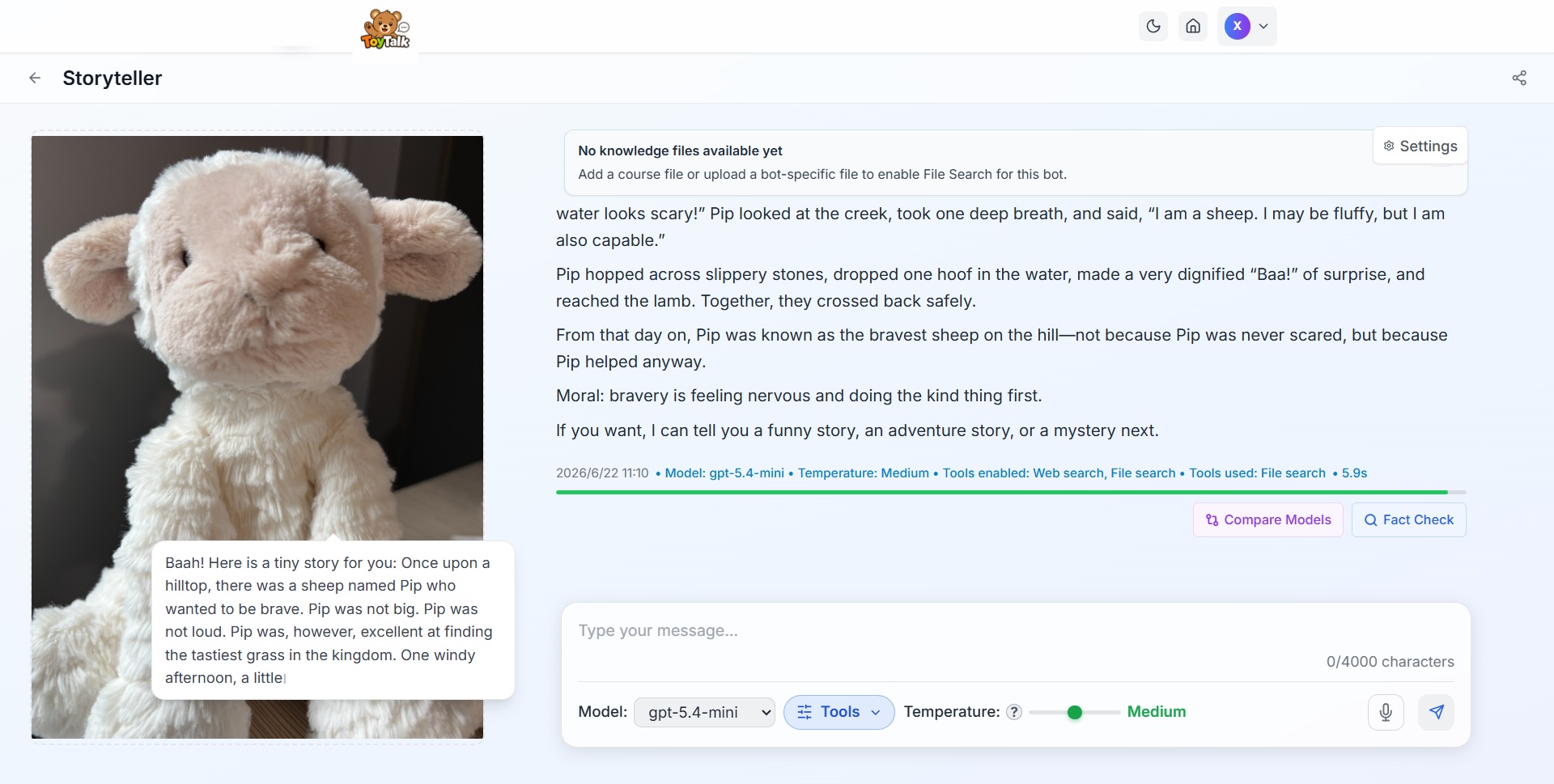}
    \caption{Screenshot of ToyTalk platform}
    \Description{Screenshot of student facing Toytalk platform page}
    \label{fig:toytalk_screenshot}
\end{figure}

We tried to address this gap through a classroom-based prototyping case study of an AI literacy and English Language Arts (ELA) curriculum built around ToyTalk (Figure \ref{fig:toytalk_screenshot}), a conversational AI toy that elementary students can customize, developed by the research team. ToyTalk allows students to create a chatbot based on their own physical toy pictures in which they can customize its purpose, characteristics, personality, communication tone and rules and guidelines. Three elementary teachers implemented the curriculum in Summer 2026 after participating in a series of professional development sessions. The research questions that guide this study are:
\begin{itemize}
    \item RQ1: How do teachers adapt their role as they implement ELA+AI literacy curriculum?
    \item RQ2: What tensions do teachers navigate when facilitating AI-powered tool mediated lesson plans?
\end{itemize}

The results of this study are based on the data from the design and implementation of the three week ELA+AI summer camp for 2nd and 3rd grade students in a rural school district in eastern part of United States. 

This paper makes contribution to the computing education community as an empirical study of teachers implementation of AI curricula and conversational toys in elementary classrooms \cite{yim_artificial_2025}. And also by specifically outlining the design implications for future iterations of student-facing AI tools for teacher improvisation in elementary classroom contexts. 

\section{Related work}
This study builds on teachers situated practice, improvisations, noticing and AI literacy for elementary students. 

\subsection{Teachers' situated practice, improvisation, and noticing}

Teachers improvise within and against their lesson plans, a phenomenon \citet{sawyer_creative_2004}
calls "disciplined improvisation", and they do so by attending and interpreting classroom cues that prompt a decision \cite{van_es_learning_2002}. This line of work frames the mismatch between prescribed materials and lived practice as a set of of enduring dilemmas teachers manage everyday \cite{lampert_how_1985}. In a qualitative study involving 26 K-12 teachers, \citet{yin_when_2026} investigated how teachers engage in adaptive practices in AI integrated classrooms. Drawing on eight focus groups data, they argue that teachers become "everyday repairers" \citet[p. ~2217]{yin_when_2026} of the values, meanings and relationships that AI systems disrupt and they call for treating relational, care-centered pedagogy as a core ethical concern that responsible AI frameworks overlook. Because this repair is relational and situated, equipping teachers to do it is crucial part of professional learning.

Research on teacher professional development (PD) emphasizes that effective PD is sustained, active and collaborative \cite{darling-hammond_effective_2017}. When teachers implement a new tool or curriculum, this management happens over time and often collectively. Teachers make sense of mandated approaches together, reshaping through their professional communities. Yet how teachers improvise and make sense of an AI-integrated curriculum with student-facing AI tool and how their role evolves through implementation remains underexplored. Han et al. \citet{han_teachers_2024} conducted a workshop with 12 families (parent-child, ages 8-12) and interviews with 16 educators to understand their perceptions on integrating generative AI into elementary writing education. Their results showed that hopes and concerns around creativity, trust, and age-appropriateness are paramount. They proposed what designers should build with oversight, role allocation, and customization of AI tools for agency. \citet{harvey_dont_2025} interviewed 29 educators and Ed-tech professionals to construct an explicitly educator centered account of the harms LLMs can introduce in classrooms, arguing that teachers' perspectives are too often absent from how such harms are framed. \citet{yoo_how_2025} interviewed K-12 teachers creating pedagogical chatbots, documenting their practices and the barriers teachers encounter such as technical issues, These studies largely capture teachers' perceptions and concerns typically through one-time interviews, and using general purpose AI platforms and few studies follow teachers as they implement specific AI-integrated curriculum with student-facing tools in elementary school context over a period of time. 
Therefore, in this study, we aim to show that how proposed design ideals, student-facing AI chatbot tool and the implementation of AI-integrated curriculum play out through teacher adaptive practices over time. 

\subsection{Chatbots and AI literacy for elementary students}

Research on AI education in elementary school shows that although students have ideas about AI, they need help to understand the black box nature of it \cite{ottenbreit-leftwich_lessons_2023}. \citet{ottenbreit-leftwich_lessons_2023} conducted a two-phased qualitative study with students and then teachers with the aim of understanding possible entry points for teaching elementary level students about AI. They conducted interviews with 10 fourth grade students and 3 elementary teachers about students' and teachers' understandings of and experiences with AI. They found that students had vague and misinformed conceptions of AI. They also found that while teachers are surprised with students' AI knowledge, they also acknowledge the challenges they have about teaching AI due to limited content knowledge about computer science and computational thinking \cite{ottenbreit-leftwich_lessons_2023}. This raises the need for involving teachers in the co-design process of elementary AI curriculum, designing AI curriculum based on students' prior knowledge, and using relatable AI concepts as entry points to students and teachers learning. 

Students increasingly are interacting with conversational agents and research in HCI has examined how young children perceive, trust, and build mental models of such agents \cite{druga_hey_2017}, including evidence that dialog agents can support early literacy outcomes such as story comprehension \cite{xu_dialogue_2022}. Recent studies have developed conversational AI tools with curricula where youth learn about AI by creating conversational agents including development environments for youth to build their own agents \cite{tian_amby_2023}, conversational-AI summer camp curricula for middle school learners \cite{katuka_summer_2023, song_ai_2023}, and broader efforts to scale AI literacy through K-12 outreach \cite{tian_ai_2026}. \citet{yang_pair-up_2023} reported a classroom-based prototyping study of a co-orchestration ecosystem in which an AI recommends student pairings and teachers make the final call. Their results showed a tension between teachers' and students' preferred levels of control over classroom transitions. More recently, studies of teacher-chatbot interaction in block-based programming examine how teachers work alongside an AI assistant in the classroom \cite{riahi_exploring_2026}, and work on LLM-based tutoring probes the promise and limits of such systems for instruction \cite{tithi_promise_2025}. 

Across these works, the analytic focus is mainly on the students' experience and learning or on the AI tools' capabilities. The teacher's mediating role around student-facing AI remains on the margin. This gap in the literature on how elementary teachers adapt an AI-integrated curriculum and a student-facing AI tools especially with elementary students and how their role evolves over time is underexplored.  Therefore, this study aims to address this gap through exploring teachers adaptations when AI technologies are implemented in real classrooms. 

\section{Method}
This research employed a qualitative case study \cite{yin_case_2009} to explore how elementary teachers make sense of and adapt their practice while implementing an ELA+AI-literacy curriculum centered on a chatbot (ToyTalk) platform. The case study approach has a scientific advantage in that it can lead to new sense making, which can then serve as a foundation for hypotheses to be tested in following studies. As this is an exploratory qualitative study, it also allows the researchers to gain insights into the context.

\subsection{Study context}

This study was conducted in a single school district in the eastern part of the United States. It took place during the Summer of 2026 in a 3-week summer camp in which 78 2nd and 3rd grade students from a rural school district learned foundational ELA and AI concepts through a curriculum that paired teacher-led sessions with ToyTalk. ToyTalk (Figure 1)is a custom-built AI chatbot platform developed by the research team to support AI literacy for elementary learners. It is a platform that enables  learners to design and build their own conversational AI toys through a no-code interface.

The ELA+AI summer camp took place as part of a reading remediation program for students that are below their grade skill levels. The curriculum, co-designed with computer science and learning science experts, aims to provide teachers with daily lesson plans, worksheets, and activities. The curriculum aims to develop foundational AI literacy for elementary students by embedding AI concepts within familiar ELA learning activities. Over the three weeks, the teachers facilitated classroom sessions where students progressed from understanding what AI is, how it generates responses to creating, training one's own AI chatbot toy, to evaluating whether the chatbot's responses are accurate or trustworthy. Each week built on the previous week's artifacts so that students could use familiar ELA practices, such as description, storytelling, reading comprehension, and discussion, to understand AI concepts.  A research team including educators, platform developers, computer scientists, and project coordinators supported implementation throughout the three weeks in June 2026.

\subsection{Participants}

Three elementary teachers (Table 1) participated. All names and identifying details have been removed and teacher quotes are attributed to anonymized teacher IDs.  All three teachers that implemented the curriculum were non-CS teachers with limited prior experience in integrating AI concepts in their teaching. They taught students from six elementary schools in a rural district which constitutes predominantly underserved and underrepresented student populations during Summer 2026. District level student demographics show 36.5\% white, 18.6\% Black, 1.6\% Asian or Asian/Pacific Islander, 36.7\% Hispanic/Latino, and 0.3\% other.

\begin{table}[ht]
  \caption{Participant demographics}
    \centering
    \begin{tabularx}{\linewidth}{@{} c c c >{\centering\arraybackslash}X >{\centering\arraybackslash}X@{}}
    \toprule
    Teacher ID & Grade & Subject & Years of experience & Prior AI experience \\
    \midrule
    T01 & 1st & Science & 14 years &  PD in 2026\\
    T02 & 4th & Mathematics & 4 years &  PD in 2026\\
    T03 & 5th & Science & 16 years &  PD in 2026\\
  \bottomrule
    \end{tabularx}
    \label{tab:part}
\end{table}

\subsection{Data Collection}

To capture teachers' experience while they reflect in action and on action \cite{schon_reflective_2016}, we drew on three  data sources collected during the camp:
\begin{itemize}
    \item Daily written individual reflections: After each camp day, each teacher completed an individual reflection using Google form created by the project team (Appendix A.2). These capture in-the-moment reasoning for each day of the camp.  These resulted in 3 documents in total with each teacher having on average 8-12 entries. 
    \item PD debrief sessions (group reflections): The group collectively reflected on each day over Zoom for all days of the camp. These sessions were 90 minutes long and were audio recorded and transcribed using Zoom. These sessions were based on each day data from the individual reflections where researchers ask and discuss on what happened during the camp class sessions. This is also the time where teachers brought their questions or concerns regarding implementation of the curriculum or students using the ToyTalk platform. These sessions resulted 9 transcript documents in total.
    \item Post-camp interview: After the summer camp ended, semi-structured individual interviews with each teacher (45 to 60 minutes long) were conducted. These were probing prior experience, specific moments of the lesson plans, tensions between the plan and student needs, and their suggestions to improve future implementation and PD needs. The interview protocol can be found in Appendix A.1. The interviews resulted in 3 transcript documents. 
\end{itemize}
Together, these sources enriched the findings through individual reflection, collective reflection, and retrospective reflection (semi-structured interview) accounts of the camp experience.

\subsection{Data Analysis}
A thematic analysis \cite{cooper_thematic_2012} approach was employed to inductively identify and code interviews, and both reflections. Due to the considerable variations of the data sources, the unit of analysis was decided on meaning. The first author worked to initially convert the individual reflections into a Microsoft Word document for each teacher. Then the first author coded each data source inductively and recorded codes and quotes in a spreadsheet file. \textit{Curriculum design and pedagogical adaptations, differentiations, technical issues with platform, recommendations for future implementations, etc}. were identified from the individual reflections data source. Then these topics were applied to code the interview and group reflection documents which led to additional codes to emerge. A total of 13 codes were identified among the three data sources. Following the inductive approach, the first author presented the emerging patterns and themes to the research team and discussed it collaboratively. The research team categorized 7 themes (initial challenges teachers faced, pedagogical adaptations and strategies, evolving understanding of AI and teacher role, technical issues and their impact, addressing diverse students needs, curriculum design, and teacher preparedness and support).

\section{Findings}

\subsection{RQ1: How do teachers adapt their role as they implement ELA+AI literacy curriculum?}

Elementary teachers adapt their roles in AI-driven curricula by embracing flexibility, modifying lesson plans, and developing new pedagogical strategies to meet their students’ needs. 

\subsubsection{Initial challenges}
Initially, teachers mentioned in their individual reflections and group debriefing reflections about facing difficulties with student engagement, technical issues with Toytalk, and curriculum pacing particularly with the low-performing students.  Technical issues with the ToyTalk platform such as microphone errors and system outages demanded teacher improvisations during classroom moments. Some of these issues were used to improve the system and some others were out of the control of the design team (e.g. microphone errors, not working computers, etc.) Some of the issues led to different iterations of the platform to improve based on teachers’ participation and feedback in the design process. 

Initially teachers reported that students struggled with crafting effective prompts, often leading to generic or incorrect AI responses. 
For example T03 mentioned that: 
\begin{quote}
    “I kind of walked behind the kids and read their questions once they’d either input them ot type them. And one of them had put, how long does a caterpillar stay in a cocoon? And so we talked about how, you know, we’re asking our toy about the story, and the toy gave him, like some example, or like, more general information, like it didn’t reference the story, so we talked about how maybe if we put in more specific, like how did the caterpillar in the story..how long did the caterpillar in the story stay in a cocoon.” (T03, June 8 group reflections)
\end{quote}

After this comment was made from the teacher, to simplify the process, the design and development team introduced pre-filled templates. These templates automatically populated fields like “role”, “job”, “communication tones”, and “rules”. This improved the AI responses making them consistent with the intended topic even when students asked off-topic questions. This led to the understanding among students that humans control AI behavior through rules and guidelines. 

\begin{figure}[htbp]
    \centering
    \includegraphics[width=\linewidth]{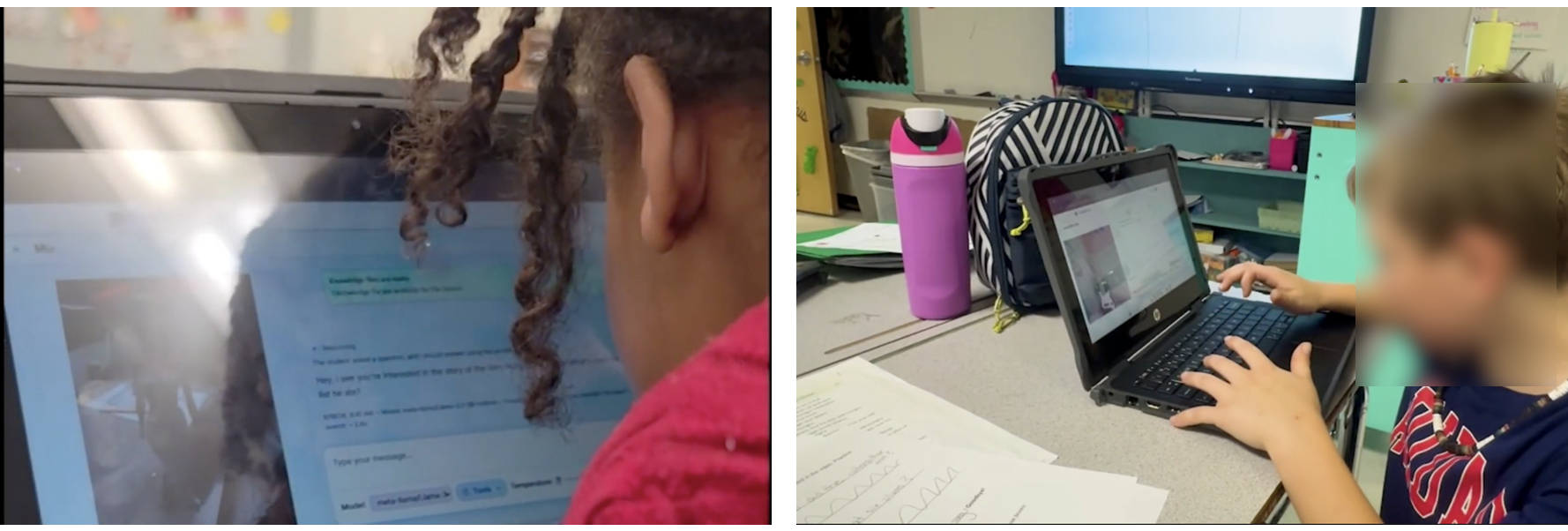}
    \caption{Teachers' point of view of students interact with ToyTalk}
    \Description{Two pictures of on the right side a blurred face student typing on his computer and on the left side a student with their hair out on the side looking at the screen on their computer attentively}
    \label{fig:teachers_pov}
\end{figure}

Teachers also reported problems with the microphone not consistently picking up students' voices. This was particularly problematic for students with speech impediments or different accents. To mitigate this, the team proposed delayed voice detection window extending the microphone’s voice detection time, and manual start/stop button which was favored by teachers as it is aligned with students’ existing familiarity with speech to text features. This change was expected to lead to less frustrating voice input interactions. These iterative changes in technical adaptations created more accessible and engaging environment within the ToyTalk platform leading to more meaningful interactions. 

In terms of curriculum pacing, teachers frequently split lessons across multiple days, especially for 2nd graders, to allow for deeper exploration and accommodate varying learning paces. For example T02 and T03 reflect on  their pacing modifications for the 2nd and 3rd grades:
\begin{quote}
     “Like, this lesson, I mean, we try to fit it in a day. Maybe taking this lesson for second grader and breaking it into two, so you can focus more and talk more about….I felt it was rushed for a second..even with only…I only had two second graders today, even with just two, because getting them to type it or say the things in there. Just I don’t think the lesson was too much for them, but maybe too much to do in one day.” (T03, June 10 group reflection)
     “We didn’t even get to finish the AI Detective today for [school 1]. The third grade was done, the second grade was not, so it’s like they got done with the first page, and it kind of all goes together…” (T02, June 11 group reflection)
\end{quote}

\subsubsection{Pedagogical adaptations and strategies}
In response to the challenges, teachers discussed in the group reflections and follow-up interviews a range of pedagogical strategies and adaptations, such as pacing adjustments, differentiated instruction and support, hands-on activities, scaffolding and modeling, and incorporating real-world scenarios. \\

\textbf{Pacing adjustments:} From the group reflection data, teachers prioritize student engagement and understanding over strict adherence to the lesson plan's timeline. They frequently split lessons across multiple days, especially for second graders to allow for deeper exploration and accommodate varying learning paces. 
The interview with the teachers showed that project team’s ability to acknowledge this need and assuring teachers that catching up was not a primary concern created a sense of comfort among teachers. 

\textbf{Differentiated instruction:} Given the diverse student needs, teachers implemented differentiated support. For example, for students struggling with writing, they facilitated whole-group discussions with the teacher transcribing responses, or proposed for alternative forms of expression like drawing or selecting emojis. They also encouraged and facilitated peer to peer learning with older (3rd graders) and more proficient students assisting their younger or struggling classmates.

\textbf{Hands-on activities:} Teachers integrated hands-on activities to enhance engagement and understanding. The creation of pattern recognition bracelets and the "Andy's Coming" brain break were some successful examples of improvisation that made abstract concepts more tangible and enjoyable to students. Teachers also mentioned movement-based activities that were included in the curriculum were found to increase student engagement.

\textbf{Scaffolding and modeling:} Teachers emphasized explicit modeling and scaffolding such as through demonstrating steps, providing sentence starters, and following the "I do, we do, you do" approach. This approach was suggested as a way to gradually release responsibility to students in forming prompts to their Toytalk toys. 
\begin{quote}
     “We say, okay , like an I do, we do, you do, so I do a question, now let's create a question together, now you create your own question. So that way, they're still forming their own question, but it gets their mind working and thinking.” (T02, June 15 group reflection)
\end{quote}

\textbf{Leveraging real-world examples:} To make AI concepts more relatable, teachers drew parallels to students' everyday experiences. Discussions about AI hallucinations were linked to familiar stories like "The Boy Who Cried Wolf". 
Teachers also used instances of technical errors as teachable moments, explaining the underlying concepts of coding and programming.  
\begin{quote}
     “We had to really explain, but I said, you know, have you ever heard the story of the boy who cried wolf, and we talked about how he was lying, but AI might not necessarily be lying, but it might give you an answer that's not correct, and if it continues to do that, you're not going to be able to trust it.” (T03, June 15 group reflection)
\end{quote}
 
Over the three weeks camp, the team learned to improvise, integrate hands-on activities, and differentiate instruction to enhance students’ learning. This led to a more responsive, student-centered approach, emphasizing patience and creativity. Teachers also recognized the importance of drawing parallels to real-world scenarios and familiar narratives in application of AI concepts. This iterative process of teaching, reflecting and adjusting highlights the required evolutions in instructional practices within AI-powered curricula. 

\subsubsection{Evolving understanding of AI and Teacher Role} Teachers' understanding of AI and their role in teaching it evolved in various ways throughout the camp. \\
\textbf{From Skepticism to understanding}: Initially, students and some teachers viewed AI as "fake" or unreliable. However, through hands-on interaction and guided exploration, they began to grasp that AI's behavior is governed by human-defined rules and guidelines. This shift in understanding was a key "aha moment" for many students. 

\textbf{Recognizing the importance of foundational skills }: The summer camp experience highlighted the critical role of basic computer literacy and reading comprehension in effectively engaging with AI tools. Teachers observed that students' struggles with typing, reading handwriting, and understanding instructions significantly impacted their ability to interact with the AI curriculum. Although the camp was designed for students who need reading remediation program, prior foundational skills were found to be important to integrate AI concepts with ELA curriculum.

\textbf{Connecting AI to broader concepts }: Teachers facilitated connections between AI concepts and broader educational themes, such as trustworthiness, critical thinking, and career exploration. Students began to think that AI systems, like humans, can make mistakes and that critical evaluation is necessary. The exposure to AI also sparked interest in programming and technology-related careers among some students. 
\begin{quote}
     “We researched coding, and I showed them what actual coding looks like, and I talked about how my husband's in IT, and how he has to do coding, and it's very difficult, but the people on the background, I said programmers and engineers, they're working on the coding. And they were like what's that look like? And I showed them and they're like wow that just looks like someone just banged the keyboard.” (T02, June 15 group reflection)
\end{quote}

Teachers facilitation of ELA+AI curriculum with elementary students is characterized by dynamic process of adaptation, improvisation, and continuous learning. While initial challenges related to students' readiness and technical issues were prevalent, teachers utilized flexibility in modifying their instructional approaches. Their evolving role emphasized patience, creativity, and a student-centered mindset that fostered a meaningful understanding of ELA and AI concepts. Teachers' experience in this study highlighted the potential of AI-integrated curriculum could ignite student curiosity, creativity, and critical thinking skills.

\subsection{RQ2: What tensions do teachers navigate when facilitating AI-powered tool mediated lesson plans?}
Teachers facilitating AI-powered tool mediated lesson plans navigate a multifaceted range of tensions. These primarily revolve around technological reliability, diverse student needs, and curriculum adaptability. These tensions necessitate significant improvisations and pedagogical adjustments. 
\subsubsection{Technical issues and their impact}
Teachers reported several technical glitches, including microphone malfunctions, slow platform response times, and various error messages. For instance, one of the teachers noted persistent microphone issues where setting indicated "allow" but the mics still wouldn't work for some students. 
\begin{quote}
    "There's a lot of mic issues, and I know I had messaged y'all and I know there was a response back, talking about check the settings. We checked the settings, the settings said the mic was allowed. Because I even physically allowed it, and I went and changed, I did everything. When you click on it, it will light up red for a second, and then go away." (T02, June 10 group reflection). 
\end{quote}
This led to frustration especially for students with speech impediments. 
Another teacher also observed that the speech-to-text function sometimes misinterpreted words like "pear" as "pair" highlighting the need for students to double-check AI inputs. 
\begin{quote}
    "So when they were using the speech to text, they would ask the question 'On Wednesday, he ate pears' or whatever the question was and it would auto-correct or it would write pair (like couple) and they wouldn't go back and read it, and so luckily I was walking around and I'd refer back to, hey this is like a homophone...they sound alike but they don't mean the same thing. Let's make sure that we're inputting you know, what we really want to ask it, you know, what answer it is we're looking for. It's not going to give us some trustworthy information, so I thought that's  good."
\end{quote}

\subsubsection{Addressing diverse students needs}
Teachers observed considerable differences in computer literacy, language barriers, reading comprehension, and engagement levels, particularly between second and third graders, and even among students within the same grade. 

\textbf{Computer literacy and foundational skills:} Many students, especially second graders and those identified as "low" (two or more grade levels behind) struggled with basic computer skills such as logging in, remembering passwords, typing, and understanding concepts like "address bar" or "space bar". This lack of foundational tech skills consumed instructional time and led to frustrations for both student and teachers. The absence of dedicated technology classes in elementary schools was identified as a contributing factor for this gap. 

\textbf{Reading and writing challenges:} The curriculum's reliance on reading and writing tasks posed a considerable challenge for many students. Students struggled with reading their own handwriting from previous days, comprehending instructions, and formulating questions or responses. This led to a common sentiment among students that there was "too much writing". 
\begin{quote}
    "I do think for some, it would make it [sentence starters] easier, but I also don't want to make it too, like, too easy where they can't learn to say or type anything." (T02, June 10 group reflections)
\end{quote}
The tension here was balancing the need for writing practice with preventing student disengagement and frustration. 
Teachers also reported that initial design of worksheets and exit tickets were challenging for some of the students, particularly those with limited writing skills.

\subsubsection{Curriculum design and adaptability}
The structure and pace of the curriculum presented another area of tension, especially regards to its suitability for diverse learning speeds and levels. 

\textbf{Pacing and lesson length }
Teachers felt that some lessons were "too long" or "too much to do in one day" for second graders, often requiring them to split lessons over multiple days. This was exacerbated by the need for extensive one-on-one support for struggling students. The rapid introduction of complex AI concepts like "hallucination" and "trustworthiness" sometimes went over students' heads, even with explicitly vocabulary.

\textbf{Differentiation and scaffolding} 
The curriculum's initial design, with its reliance on worksheets and independent work, proved challenging for lower-performing students. Teachers found themselves constantly improvising, simplifying instructions, and providing more scaffolding than anticipated. The suggestion to integrate "I do, we do, you do" modeling was proposed to help students, especially with question formulation. 
The idea of providing pre-filled templates or multiple-choice options for certain tasks was also discussed to ease the burden on students with limited writing skills. 

\textbf{Balancing structure and creativity} 
While the curriculum aimed to foster creativity, some teachers noted a tension between encouraging open-ended exploration and providing enough structure for students to succeed. For example, initially allowing students to design their toy's purpose, and then later directing them to focus on the "Very Hungry Caterpillar" story, created confusion and a "mismatch" for some students who had already formed a strong personal connection with their toy's initial purpose. 
\begin{quote}
    "At the beginning, they designed their toy to do their special job, you know, like to protect me, or to be a friend, and then they all designed the one to do the very hungry caterpillar story. So I think they really thought...well, what do I want mine to do. And then we told them, it's got to tell the story of the very hungry caterpillar." (T03, interview)
\end{quote}
This highlights the need for clearer alignment between creative freedom and specific learning objectives. 

\subsubsection{Teacher preparedness and support}
Teachers experienced tensions related to their own preparedness and the support systems in place. All were new to teaching AI concepts and felt they were "learning with the kids". The rapid pace of curriculum development, with materials sometimes finalized just before implementation, added to this stress. However, the collaborative briefing sessions were highly valued, providing a platform for teachers to share challenges, suggest modifications, and feel "heard". This iterative feedback loop was crucial to effectively address emerging technological, learner, or instructional tensions. 

\section{Discussion}

Our findings show that an ELA+AI-literacy curriculum delivered through the ToyTalk conversational chatbot ran through teachers' continuous adaptive work. Below we discuss the implications to the design and implementation in the field. 

\subsection{Teachers adaptive practice in AI-integrated curriculum}

Teachers sat at the intersection of three persistent tensions: technology, learners, and instruction. These are dilemmas of practice that teaching entails \cite{lampert_how_1985}. The adaptive work that teachers performed at this intersection can be organized into the following four modes. Closest prior literature of teachers orchestrating technology integrated curriculum in elementary classrooms is Peddycord-Liu et al.'s \cite{peddycord-liu_field_2019} year long field study of teachers using ST Math, which derived four teacher-orchestration activity types: preparation, integration, intervention, and data-informed practice. As our data foregrounds how teachers continually adapt a generative, emerging AI tool in a prototyping form, we re-frame orchestration activities as practices under tension and extend the framework in the four modes below. 
Two of the modes map onto \citet{peddycord-liu_field_2019} interventions and integration activities. The other two are demanded specifically by a generative, unpredictable AI tools and can be seen as peripheral extensions of the framework.   
\begin{itemize}
    \item \textbf{Repair}: troubleshooting technology issues such as microphones, switching to backups, or helping students when unexpected computer/tool use issue arises. This is a form of human-AI work in which the teacher supplies the robustness the system lacks \cite{du_boulay_human-ai_2023}. 
    \item \textbf{Differentiate}: adjusting pacing and modality to a heterogeneous class through splitting lessons across days, transcribing for struggling writers, and facilitating peer tutoring enacts responsive, situated teaching triggered by real-time noticing of student cues (e.g. reading and correcting students prompts) \cite{van_es_learning_2002}. This corresponds closely to \citet{peddycord-liu_field_2019} \textit{Intervention}, as teachers play crucial role in students' cognitive and affective learning processes. They are also managing students' progress and facilitating peer tutoring.   
    \item \textbf{Translate}: making abstract AI concepts digestible for young students such as grounding "hallucination" in The Boy Who Cried Wolf, or framing a speech to text slip as a homophone and a lesson in verifying AI output. This parallels with emerging work scaffolding children to detect AI errors \cite{tian_when_2026}. This corresponds closely to \citet{peddycord-liu_field_2019} \textit{Integration}, as it extends the technology connect to classroom content into AI topics.  
    \item \textbf{Balance}: teachers were also negotiating learning objectives with students' creative agency. for example, when students first designed their toy's purpose and then were redirected to make it tell "The Very Hungry Caterpillar" story. They were able to implement the conversational-AI curricula  that let students retain authored identity while meeting instructional goals \cite{tian_amby_2023}.
\end{itemize}

\begin{figure}[h]
  \centering
 \includegraphics[width=\linewidth]{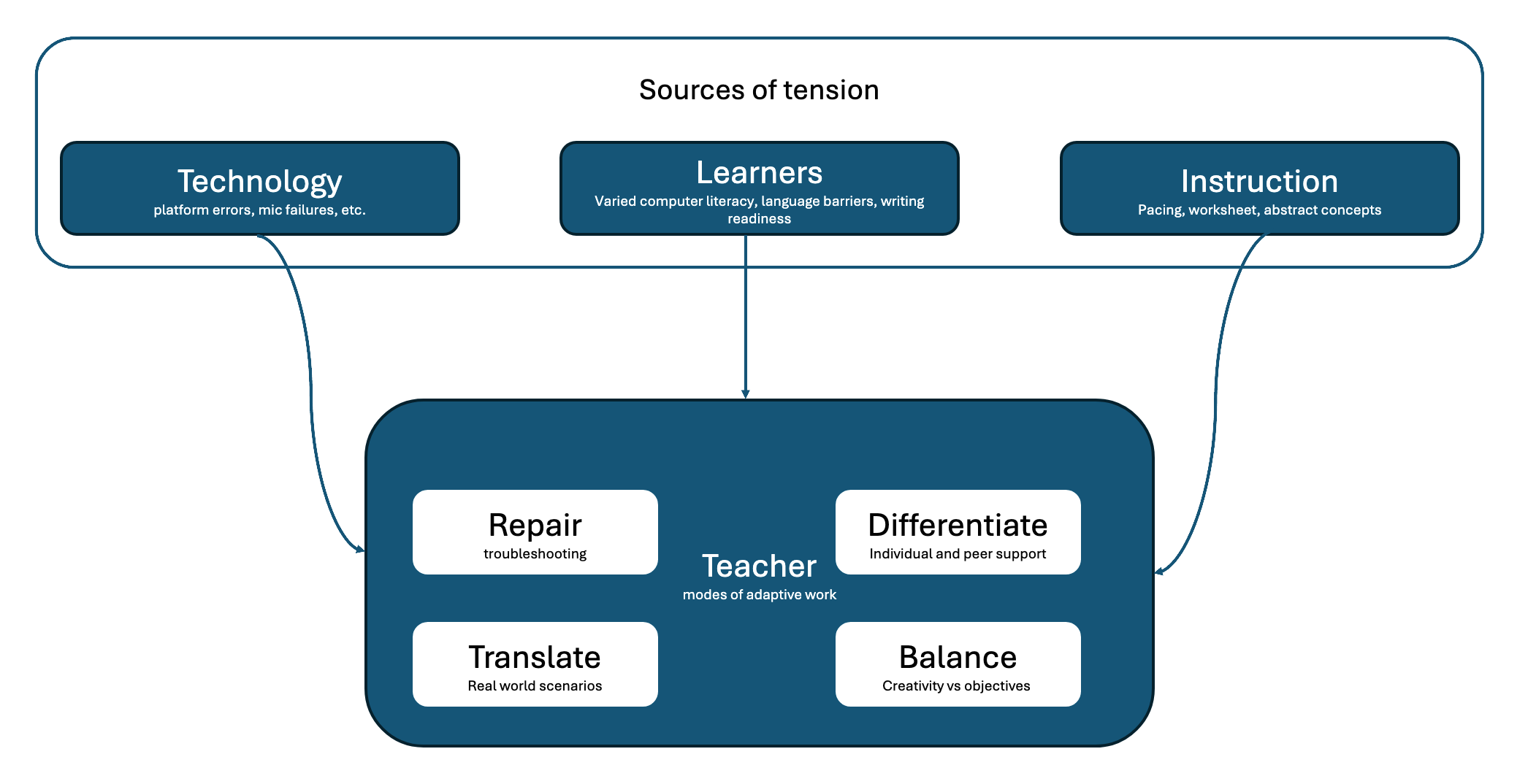}
  \caption{Teacher adaptive practices in AI-integrated elementary classrooms}
  \Description{Two overall boxes describing sources of tension and teacher adaptive work. Within the sources of tension there are boxes for technology, learners, and instruction. Within the teacher adaptive work are repair, differentiate, translate, and balance.}
\end{figure}

\subsection{Sense making as a collective}
Teachers description of "learning with the kids" and the group reflection sessions as they raised problems, proposed modifications, and felt "heard" is a collective interpretations and sense making. This mirrors accounts of teachers collectively creating a new approach through professional community \cite{dogan_artificial_2025} and aligns with features of effective professional development \cite{darling-hammond_effective_2017}. Teachers' AI-specific knowledge was emerging as they brought strong pedagogical and content expertise and built technological understanding in real time \cite{mishra_tpack_2023}. 

\subsection{Design implications}
Because adaptation is the mechanism of implementation, AI curricula and student-facing tools should be designed for teacher adaptation: 
\begin{itemize}
    \item Design for repair: designs should include the assumption of unreliable classroom technology infrastructure and provide manual options for teachers to fallback on. 
    \item Scaffold foundational skills gap: because typing, reading, and prompt-formulation limited participation for some students, building a multimodal input with sentence starters, pre-filled templates while preserving students' productive struggle in writing practice is essential. 
    \item Support conceptual translation/customization: providing teachers with ready analogies, worked examples and prompts is necessary to implement AI curriculum. 
    \item implement design-iteration-reflection loop: teachers need structured channels (daily individual, group reflections) as part of the system implementation. 
\end{itemize}

Teachers' constant navigation of tensions and their adaptive labor bridge the gap for the successful implementation of the ToyTalk tool and the ELA+AI literacy curriculum. This means AI tools and curricula for elementary level should be designed for teacher improvisation.

\subsection{Conclusion}
We studied how elementary teachers implemented an ELA+AI literacy curriculum built around ToyTalk, a conversational AI toy. Over three week summer camp with elementary students, we identified that teachers repaired failing technology, differentiated for gaps in learners' readiness, translated abstract AI concepts through familiar stories, and balanced children's creative agency with learning goals. Over the three weeks teachers help to responsively redesign the curriculum to fit their students' needs. 

From these findings, we developed a framework in which teachers sit at the intersection of three tensions (technology, learners, and instruction) and perform four modes of adaptive practices (repair, differentiate, translate and balance). This framework extends prior accounts of teacher orchestration of technology-integrated curriculum \cite{peddycord-liu_field_2019, kangas_qualitative_2017} and surfacing modes that generative AI tools demand. 

This was a classroom-based prototyping study based on one three-week camp with a small number of teachers and a student population enrolled in reading-remediation program. Findings should be taken as snapshots and center teachers' accounts rather than objective classroom measures. Future work may follow the design iteration across multiple sites and full school-year implementation, incorporating students' perspectives and test whether the four modes of adaptive practice applies to other AI curricula and tools.

\newpage
\begin{acks}
This work was supported by National Science Foundation Grants DRL-2524505. Findings, and conclusions expressed in this paper are those of the authors and do not necessarily reflect the views of the National Science Foundation. We are grateful to the elementary teachers who have worked with us as part of this project.
\end{acks}

\textbf{Ethical Approval:}
The study presented in this paper was approved by the NC State University's Internal Review Board number 29017. 

\bibliographystyle{ACM-Reference-Format}
\newpage
\bibliography{references}

\newpage
\appendix

\section{Appendices}

\subsection{Interview Protocol}

\begin{itemize}
    \item Before this camp, what was your experience with AI tools in the classroom? 
    \item Think about a moment during the camp (e.g., specific details from the reflection responses) when you made a decision that wasn’t in the plan. Walk me through it.
        \item What were you noticing in the room that made you decide to do that?
        \item What knowledge or experience were you drawing on when you made that call? Any additional moments you want to talk about? 
        \item What do you think physical or embodied experiences do that screen-based activities won’t be able to achieve?
        \item What hands-on components would you add if you were designing the AI lessons for the next camp?
    \item Were there times you felt tension between following the lesson plan and doing what students needed?
    \item Is there anything that happened during the camp that has been sitting with you and that you would like to share?
    \item Looking across the entire 3-week curriculum, what do you see as its biggest strengths? what parts of the curriculum went particularly well and what parts/lessons can be improved? 
    \item If you were to implement this during the school year, what would need to change?
        \item What would a realistic implementation look like (timing, duration, integration with existing curriculum)?
        \item If you are preparing other teachers to lead this activity, how would you do the PD? What support would teachers need to successfully adopt this curriculum?

\end{itemize}

\subsection{Daily reflection survey prompts }
\begin{itemize}
    \item What is the most important thing about AI that was explored today
    \item Please write down what went well today
    \item What could be improved
    \item What do you think is most important to remember about how today's lesson went
    \item Anything inspiring to report?
    \item What AI toys did the students make today? Are there any creations that surprised you or stood out to the class?

\end{itemize}

\end{document}